\documentclass[aps,prb,reprint,superscriptaddress,nofootinbib]{revtex4-2}

\usepackage[T1]{fontenc}
\usepackage[utf8]{inputenc}
\usepackage[english]{babel}
\usepackage{graphicx}
\usepackage{amsmath,amssymb,bm}
\usepackage{booktabs}
\usepackage[colorlinks=true,linkcolor=blue,citecolor=blue,urlcolor=blue]{hyperref}
\usepackage{placeins}
\begin{document}

\title{Robustness of the Verwey transition against remanent strain-induced defects in magnetite}

\author{M.~A. Gala}
\affiliation{AGH University of Krakow, Faculty of Physics and Applied Computer Science, 30-059 Krak\'ow, Poland}
\author{K.~J. Grzywa}
\affiliation{AGH University of Krakow, Faculty of Physics and Applied Computer Science, 30-059 Krak\'ow, Poland}
\author{K.~Kom\k{e}dera}
\affiliation{AGH University of Krakow, Faculty of Physics and Applied Computer Science, 30-059 Krak\'ow, Poland}
\author{R. Zalecki}
\affiliation{AGH University of Krakow, Faculty of Physics and Applied Computer Science, 30-059 Krak\'ow, Poland}
\author{A. Baczma\'nski}
\affiliation{AGH University of Krakow, Faculty of Physics and Applied Computer Science, 30-059 Krak\'ow, Poland}
\author{Z. K\k{a}kol}
\affiliation{AGH University of Krakow, Faculty of Physics and Applied Computer Science, 30-059 Krak\'ow, Poland}
\author{A.~Pacanowska}
\affiliation{Institute of Nuclear Physics, Polish Academy of Sciences, 31-342 Krak\'ow, Poland}
\author{C.~M.~N. Kumar}
\affiliation{Institute of Nuclear Physics, Polish Academy of Sciences, 31-342 Krak\'ow, Poland}
\author{R. Rodriguez Lamas}
\affiliation{European Synchrotron Radiation Facility, 71 Avenue des Martyrs, CS 40220, 38043 Grenoble Cedex 9, France}
\author{C. Detlefs}
\affiliation{European Synchrotron Radiation Facility, 71 Avenue des Martyrs, CS 40220, 38043 Grenoble Cedex 9, France}
\author{C. Yildirim}
\affiliation{European Synchrotron Radiation Facility, 71 Avenue des Martyrs, CS 40220, 38043 Grenoble Cedex 9, France}
\author{C. Marin}
\affiliation{Univ. Grenoble Alpes, Grenoble INP, CEA, IRIG, PHELIQS, 38000 Grenoble, France}
\author{J.~Debray}
\affiliation{Institut N\'eel, CNRS and Univ. Grenoble Alpes, 38042 Grenoble, France}
\author{G. Beutier}
\affiliation{Univ. Grenoble Alpes, CNRS, Grenoble INP, SIMaP, 38000 Grenoble, France}
\author{L. Ortega}
\affiliation{Laboratoire de Physique des Solides, Universit\'e Paris-Saclay, Campus d'Orsay, 91405 Orsay, France}
\author{A.~Koz\l{}owski}
\affiliation{AGH University of Krakow, Faculty of Physics and Applied Computer Science, 30-059 Krak\'ow, Poland}
\author{N.~Bari\v{s}i\'c}
\affiliation{Institute of Solid State Physics, TU Wien, 1040 Vienna, Austria}
\affiliation{Department of Physics, Faculty of Science, University of Zagreb, HR-10000 Zagreb, Croatia}
\author{J.~E. Lorenzo}
\affiliation{Institut N\'eel, CNRS and Univ. Grenoble Alpes, 38042 Grenoble, France}
\author{W.~Tabi\'s}
\email{wtabis@agh.edu.pl}
\affiliation{AGH University of Krakow, Faculty of Physics and Applied Computer Science, 30-059 Krak\'ow, Poland}

\date{\today}

\begin{abstract}
The Verwey transition in magnetite is a benchmark electronically driven phase transition that is highly sensitive to lattice imperfections and external perturbations. Doping, deviations from oxygen stoichiometry, and irradiation-induced point defects are known to lower the Verwey transition temperature, $T_{\rm V}$. By contrast, the role of remanent strain fields and extended defect structures generated by uniaxial stress has remained largely unexplored. Here we combine dark-field x-ray microscopy (DFXM) and ac magnetic susceptibility to determine how strain-induced defects affect the Verwey transition in stoichiometric single-crystalline magnetite after uniaxial compression and unloading. The DFXM measurements were performed on two samples compressed along the same $[011]$ direction: loading to $200~\mathrm{MPa}$ generated stable line-like defects and remanent strain fields, whereas loading beyond the fracture threshold produced denser defect networks and mechanical fracture. To test the effect of a different loading geometry, we also measured the ac susceptibility of a sample fractured by compression along $[001]$. Real-space DFXM imaging, lattice-orientation mapping, and residual strain mapping show substantial remanent structural disorder after compression. Nevertheless, neither the transition temperature $T_{\rm V}$ nor the sharpness of the transition changes measurably in any of the compressed or fractured samples. This behavior contrasts with the strong suppression of $T_{\rm V}$ by point-like disorder. Our results show that extended line-like defects and remanent strain fields produced by plastic deformation do not measurably perturb the macroscopic electronic ordering responsible for the Verwey transition, thereby distinguishing the effects of point defects from those of extended slip-related defects in magnetite.
\end{abstract}

\maketitle

\section{Introduction}

Defects are unavoidable in real crystals and can strongly modify macroscopic physical properties. Their influence is especially important near phase transitions, where small perturbations of the lattice, charge, or magnetic degrees of freedom may change the transition temperature, order, or hysteresis. Depending on their dimensionality, defects may appear as point defects, such as vacancies or dopant atoms; line defects, such as dislocations; planar defects, such as slip or twin boundaries; or extended defects, such as grain boundaries. Distinguishing the effects of these different defect classes is essential for understanding phase transitions in correlated materials.

Here we study this problem in magnetite, Fe$_3$O$_4$, by examining whether remanent lattice defects induced by uniaxial compression affect the Verwey transition. We focus on stoichiometric single crystals that were compressed at room temperature and then measured after unloading. DFXM reveals persistent lattice distortions and line-like defects produced by uniaxial compression to $200~\mathrm{MPa}$ and beyond, whereas ac susceptibility shows that the Verwey transition temperature, $T_{\rm V}$, remains unchanged, even in samples that fractured during compression.

\begin{figure*}[t]
\centering
\includegraphics[width=\textwidth]{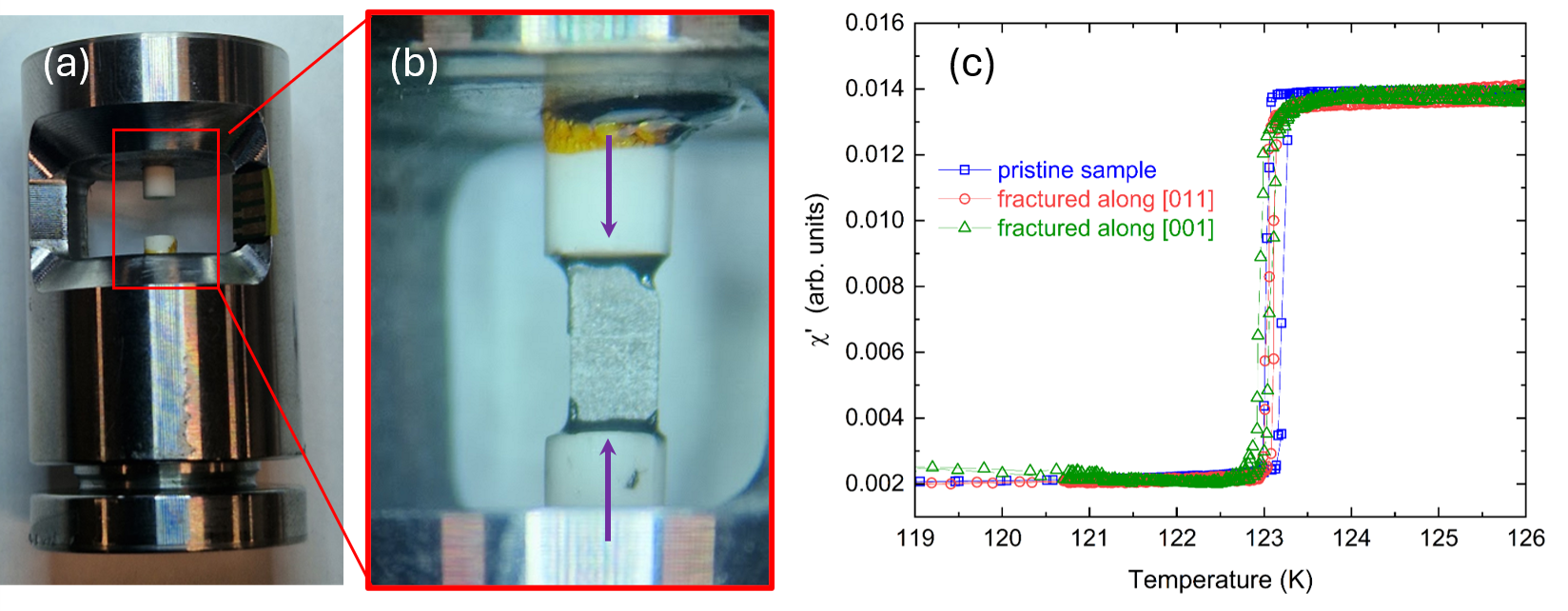}
\caption{Uniaxial compression and magnetic characterization of magnetite. (a) Compression device consisting of a cylindrical steel housing and a central piston. (b) Magnetite sample mounted between ceramic posts with Stycast epoxy; purple arrows indicate the direction of applied uniaxial compression. (c) Real part of the ac susceptibility, $\chi'$, for the pristine reference sample and for samples fractured by compression along $[011]$ and $[001]$ directions. Within experimental resolution, the Verwey transition temperature remains unchanged after compression and fracture.}
\label{fig:compression}
\end{figure*}

The Verwey transition has been studied since the early observation of low-temperature anomalies in magnetite more than a century ago~\cite{Renger1913}. It remains a prototypical example of a coupled electronic, lattice, and magnetic phase transformation~\cite{Piekarz2006}. Subsequent studies have examined this coupling from several complementary perspectives, including lattice dynamics across the transition and trimeron--phonon coupling~\cite{Kolodziej2012,Piekarz2021}, magnetic-field-induced structural and electronic rearrangements~\cite{Kolodziej2020}, temperature-driven self-doping of the octahedral Fe sublattice~\cite{Elnaggar2021}, and correlations between $T_{\rm V}$, magnetic properties, and electrical transport~\cite{Podgorska2025}. Magnetite is therefore an excellent model system for studying how external perturbations such as doping, pressure, and defects affect electronically ordered states. Deviations from oxygen stoichiometry and chemical substitution are well known to reduce $T_{\rm V}$ and to broaden the transition, changing it from a sharp discontinuous anomaly into a more continuous response. For some dopants and for sufficiently large deviations from stoichiometry, the transition becomes continuous and is eventually suppressed entirely at high doping levels~\cite{Aragon1986,Kozlowski1996,Kakol1989,Kakol1991}. Although the microscopic distinction between first- and second-order behavior has been discussed in earlier and recent work~\cite{Walz1990,Prozorov2023}, the strong sensitivity of $T_{\rm V}$ to disorder is firmly established. In particular, irradiation studies have shown that point defects alone can lower $T_{\rm V}$ while preserving a sharp transition~\cite{Prozorov2023}, raising the question of whether the dominant perturbation is carrier doping, local lattice distortion, or both.

Pressure provides a complementary way to perturb the lattice. Measurements performed under hydrostatic pressure generally show that $T_{\rm V}$ decreases while pressure is applied~\cite{Coe2012,Kolodziej2023,Pasternak2003,Kuriki2002,Klotz2006,Spalek2008}. Measurements after pressure release probe a different regime: pressure cycling into the GPa range can leave a persistent increase in $T_{\rm V}$, with the magnitude depending on the peak pressure and sample composition~\cite{Caporzen2010,Bialo2019}. This distinction is important because measurements under load probe the thermodynamic response to pressure, whereas measurements after decompression probe the remanent defect and strain state left in the crystal. The microscopic origin of this pressure memory remains unresolved, because the corresponding defect structure has not been directly imaged.

Uniaxial stress is distinct from hydrostatic pressure because it breaks the cubic symmetry and couples directly to selected crystallographic directions. Its effect on $T_{\rm V}$ is therefore strongly orientation dependent~\cite{Coe2012}. Experimentally, $T_{\rm V}$ was found to increase under uniaxial compression along the cubic $[111]$, $[110]$, and $[001]$ directions, with the strongest response reported for $[111]$ loading~\cite{Balberg1976}; an increase of $T_{\rm V}$ under compression along $[110]$ was also reported by Nagasawa \textit{et al.}~\cite{Nagasawa2007}. The remaining question is different: whether the plastic deformation and residual strain left after removal of the uniaxial load produce a measurable change in $T_{\rm V}$. This question is particularly relevant because uniaxial stress can activate crystallographic slip and create line-like defects whose spatial distribution is highly anisotropic. The residual defect structure created by such loading, and its possible effect on the Verwey transition after unloading, have not been systematically studied.

The present work addresses this gap. We apply controlled uniaxial compression to high-quality stoichiometric magnetite crystals cut from the same parent crystal, image the remanent defect structure by DFXM, and determine $T_{\rm V}$ by ac susceptibility. The results show that uniaxial stress applied along the $[011]$ direction creates persistent strain fields and line-like defects, but these extended defects do not measurably shift or broaden the Verwey transition. This contrast between the strong effect of point-like disorder and the weak effect of extended remanent defects provides a useful distinction for interpreting disorder effects in magnetite and other correlated oxides.

\section{Experimental}

Four single-crystalline magnetite samples were prepared from a larger crystal grown by the optical floating-zone method. The samples were cut into rectangular bars with typical dimensions of $1 \times 0.3 \times 0.5~\mathrm{mm}^{3}$. Three samples were oriented with the $[011]$ direction along the longest edge. One of them was kept as the pristine reference sample. The second was compressed along the same $[011]$ direction up to $200~\mathrm{MPa}$ without macroscopic fracture and is referred to below as the moderately compressed sample. The third was compressed along $[011]$ beyond $200~\mathrm{MPa}$, which produced permanent fracture, and is referred to as the fractured sample. Thus, both deformed samples investigated by DFXM were compressed along the $[011]$ direction. The uniaxial load was applied at room temperature. The three $[011]$-oriented samples were characterized by ac susceptibility and DFXM; for the two compressed samples, the measurements were performed after unloading. An additional sample, oriented with the $[001]$ direction along the longest edge, was compressed to fracture and then measured by ac susceptibility. This $[001]$-compressed sample was used to compare the fracture behavior and the ac-susceptibility response, but it became too brittle after fracture to be prepared for DFXM.

\begin{figure}[t]
\centering
\includegraphics[width=0.5\textwidth]{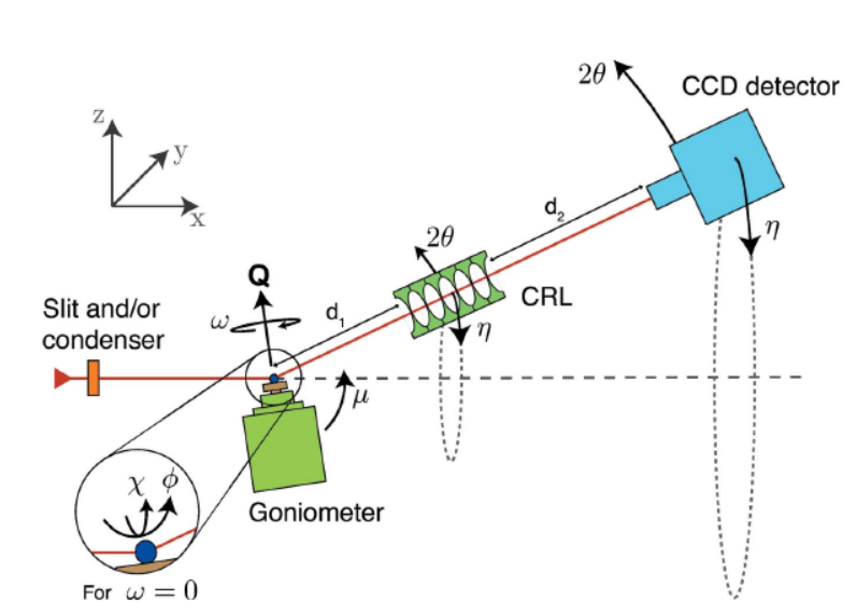}
\caption{Schematic of the DFXM setup at the ESRF ID03 beamline. A monochromatic x-ray beam illuminates the sample, which is mounted on a goniometer. A selected Bragg reflection from the cubic $\{400\}$ family is magnified by compound refractive lenses and imaged on a two-dimensional detector. Rocking and rolling scans probe lattice misorientation, while scans of $2\theta$ probe local changes in lattice spacing. The DFXM end station at ID03 was relocated and upgraded from the former ID06-HXM instrument~\cite{Isern2025}. Schematic adapted from Ref.~\cite{Poulsen2017}.}
\label{fig:beamline}
\end{figure}

Uniaxial stress was applied with the compression cell shown in Fig.~\ref{fig:compression}. After polishing, each crystal was mounted between ceramic posts using Stycast epoxy to distribute the force over the sample face. The posts were attached to a stainless-steel frame that provided controlled directional compression. Consistent with the sample description above, compression was applied along either $[011]$ or $[001]$ crystallographic directions, and in selected cases the applied stress was increased above the fracture threshold.

The Verwey transition was characterized before and after compression by measuring the ac magnetic susceptibility with a custom-built magnetometer based on a counter-wound coil assembly. A chromel--Au--0.03~at.\% Fe thermocouple from Lake Shore was placed close to the sample to monitor the temperature. Measurements were performed continuously with a temperature ramp rate not exceeding $0.1~\mathrm{K/min}$, so that possible changes in the transition temperature and transition width could be resolved accurately. The sample response was detected with an SR830 lock-in amplifier. Typical measurement parameters included a driving voltage of 5~V applied to the excitation coil at a frequency of 188.88~Hz.

\begin{figure*}
\centering
\includegraphics[width=0.7\textwidth]{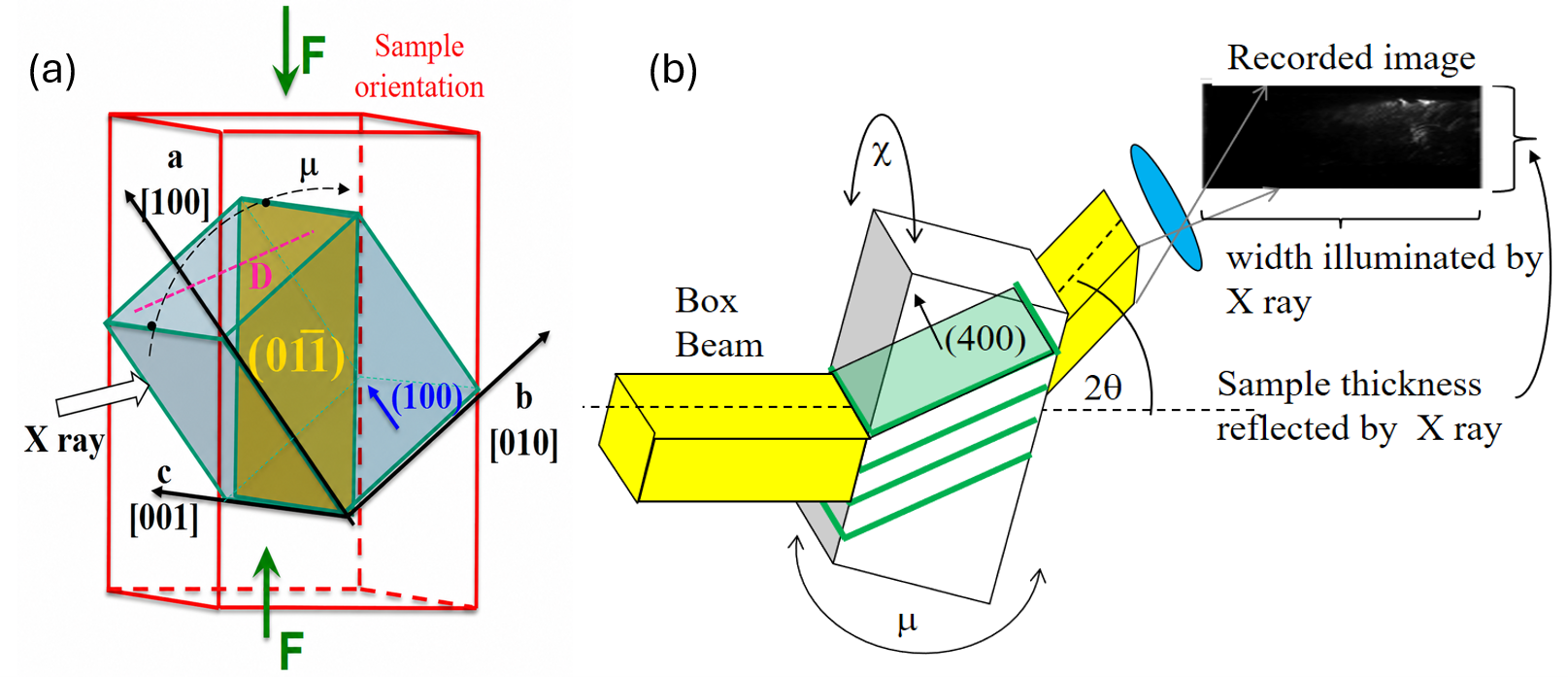}
\caption{DFXM geometry used to image the compressed magnetite crystals. (a) Sample orientation relative to the incident x-ray beam. The sample was compressed along $[011]$, with a $(01\bar{1})$ surface exposed in the DFXM geometry. The cubic $(400)$ reflection was selected for imaging. (b) Schematic of the DFXM optical layout in the box-beam geometry. The vertical direction of the recorded image corresponds to the sample thickness, and the x-ray objective inverts the image, reversing the front and rear sides of the sample.}
\label{fig:diffraction}
\end{figure*}

The remanent defect structure was investigated by DFXM at the ESRF ID03 beamline, using the DFXM end station relocated and upgraded from the former ID06-HXM beamline~\cite{Isern2025}. DFXM selects a specific Bragg reflection, magnifies the diffracted beam using compound refractive lenses, and records a real-space image of the diffracting volume. This makes it possible to image bulk lattice distortions in crystals with thicknesses of order $100~\mu\mathrm{m}$, rather than only surface damage. In the present geometry, the cubic $(400)$ reflection was selected for imaging, corresponding to a scattering angle of approximately $2\theta=20.0^\circ$ at 17~keV. Local lattice misorientation was determined from scans of the sample rotations $\mu$ and $\chi$, and local strain was determined from shifts of the Bragg-peak position in $2\theta$.
The experimental layout is shown schematically in Fig.~\ref{fig:beamline}.

The experiment used 17~keV x rays from an undulator source. The photon energy was selected with a Si(111) monochromator, giving a relative bandwidth $\Delta E/E=10^{-4}$~\cite{Yildirim2023}. Both a box-shaped beam of $100\times100~\mu\mathrm{m}^{2}$ and a line-shaped beam of $100\times1~\mu\mathrm{m}^{2}$ were tested. The data discussed below were acquired in the box-beam configuration, in which a fixed region in the central part of the sample was illuminated. After magnification by the compound-refractive-lens DFXM objective, the diffracted beam was recorded with the high-resolution far-field imaging system, consisting of a scintillator-coupled Optique Peter Twinmic optical microscope equipped with a $10\times$ objective and a pco.edge 4.2 BI back-illuminated sCMOS camera~\cite{Isern2025}. The resulting images provide micrometer-scale spatial resolution over the illuminated volume and contain contrast from local changes in the fraction of the crystal satisfying the Bragg condition.

The sample was first positioned and aligned using a retractable near-field camera located close to the sample and upstream of the DFXM objective. The quality of the selected Bragg reflection was checked during this alignment, after which the near-field camera was retracted. Small scans of the $\mu$ and $\chi$ rotation stages were used to characterize the local lattice-orientation distribution, whereas local changes in $d$ spacing were obtained by scanning $2\theta$. In the box-beam measurements, a fixed region in the central part of the sample was imaged, as shown schematically in Fig.~\ref{fig:diffraction}(b).

\section{Results}

\subsection{DFXM contrast of the remanent defect structure}

\begin{figure*}[!th]
\centering
\includegraphics[width=\textwidth]{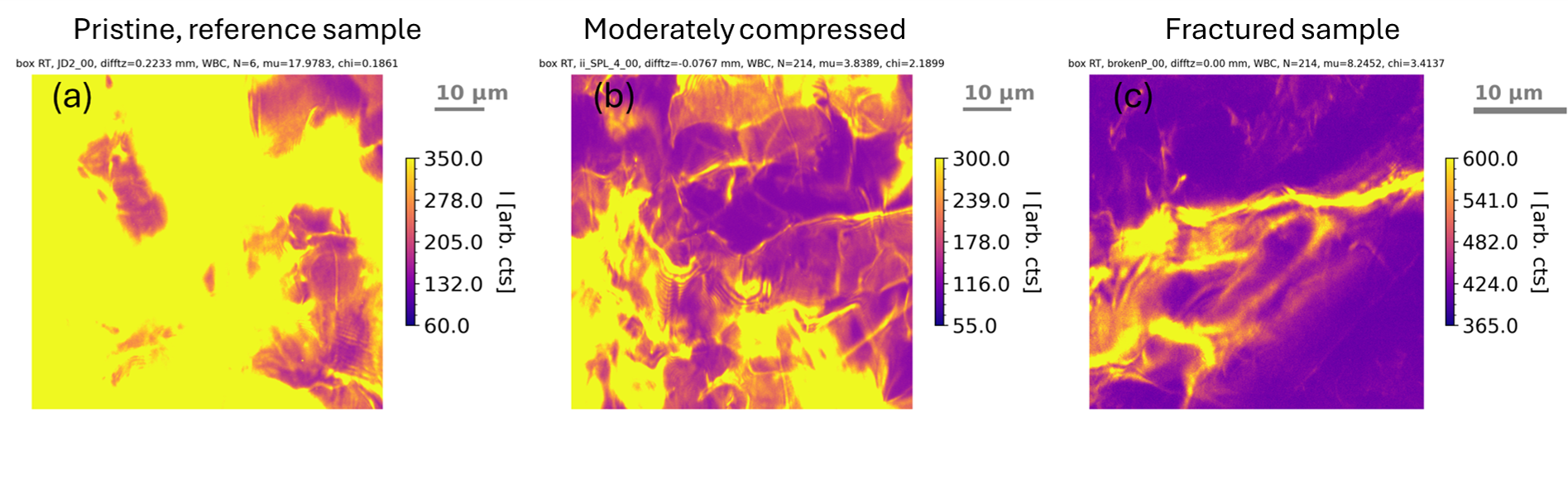}
\caption{Recorded DFXM peak images obtained at the optimal $\mu$ and $\chi$ angles of the $(400)$ reflection after different compression treatments. (a) Pristine reference sample. (b) Sample compressed to $200~\mathrm{MPa}$ along $[011]$ without fracture. (c) Sample compressed along $[011]$ beyond the fracture threshold. The progressively stronger spatial inhomogeneity of the Bragg-peak intensity from (a) to (c) reflects increasing heterogeneity of the diffracting volume after uniaxial compression; the respective contributions of lattice orientation and residual strain are quantified in Fig.~\ref{fig:XRDMF}. The color scale is adjusted independently for each panel and therefore should be used to assess contrast within a given image, not to compare absolute intensities between panels. Scale bars denote $10~\mu\mathrm{m}$.}
\label{fig:004_peak}
\end{figure*}

\begin{figure*}
\centering
\includegraphics[width=\textwidth]{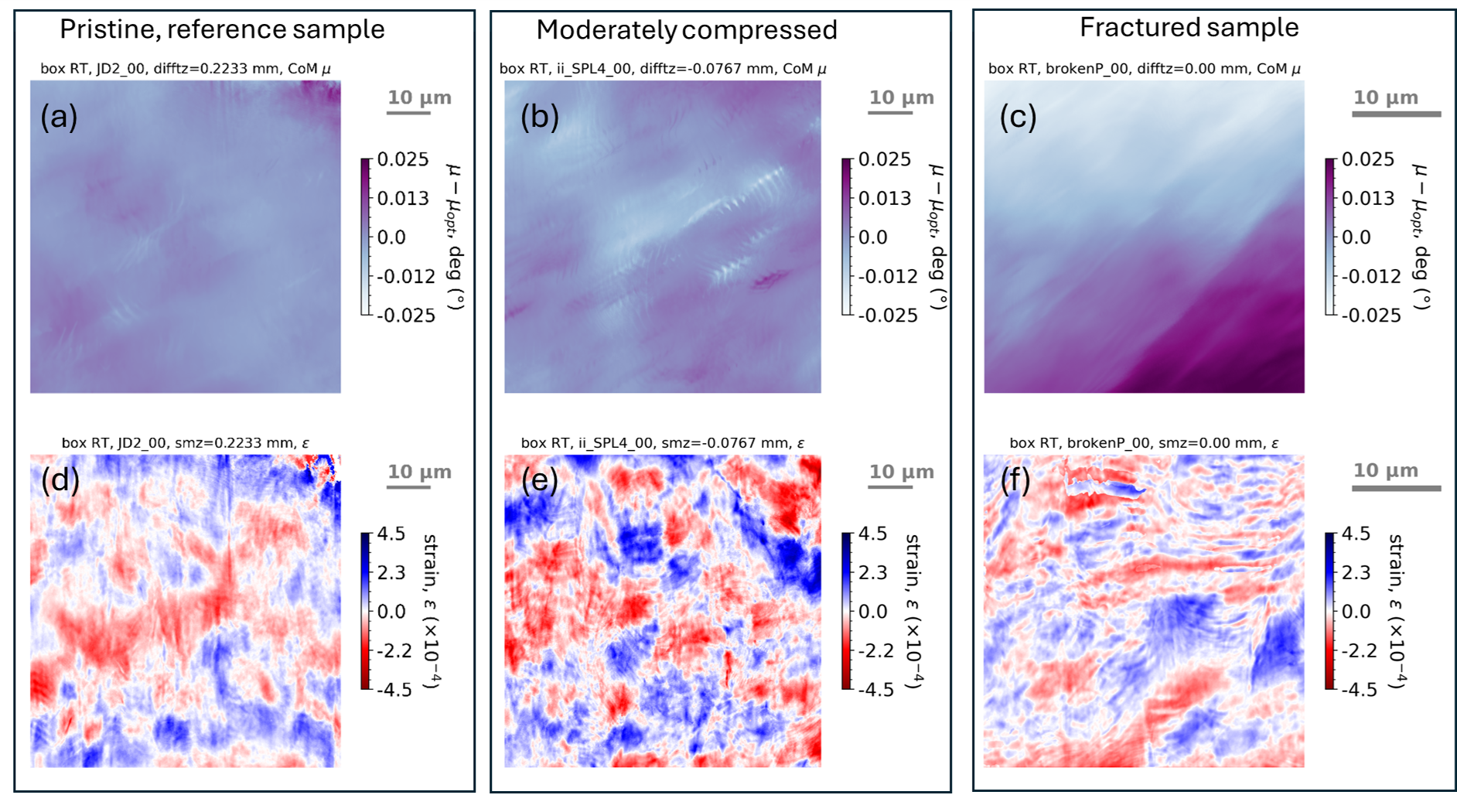}
\caption{DFXM maps of lattice misorientation and residual axial strain after compression and unloading. The measurements were performed without external stress applied during DFXM imaging. The left column shows the pristine reference sample, the middle column shows the sample compressed along $[011]$ to $200~\mathrm{MPa}$ and then unloaded without macroscopic fracture, and the right column shows the sample compressed along $[011]$ beyond the fracture threshold and then unloaded. (a)--(c) Maps of the center of mass of the local $\mu$ rocking curves. (d)--(f) Corresponding residual axial strain maps derived from shifts of $2\theta$ for the same regions of interest. Compression produces line-like and band-like misorientation contrast together with heterogeneous residual strain fields. In the fractured sample, partial strain release is accompanied by stripe-like strain contrast. Scale bars denote $10~\mu\mathrm{m}$.}
\label{fig:XRDMF}
\end{figure*}

Figure~\ref{fig:004_peak} shows detector images recorded near the maximum of the $(400)$ Bragg reflection for the pristine reference crystal, the moderately compressed crystal, and the fractured crystal compressed along $[011]$. These images are not full orientation or strain maps. Instead, they show the spatial distribution of diffracted intensity at the angular setting for which the largest part of the illuminated region satisfies the Bragg condition. They therefore provide a direct visual comparison of the homogeneity of the diffracting volume in the three samples. Quantitative separation of lattice misorientation and axial strain is provided by the maps in Fig.~\ref{fig:XRDMF}.

The pristine sample [Fig.~\ref{fig:004_peak}(a)] exhibits a largely uniform intensity distribution, consistent with a low density of extended defects and a nearly homogeneous lattice orientation within the probed volume. This image provides the structural baseline for comparison with the compressed samples. After compression to $200~\mathrm{MPa}$ without macroscopic fracture [Fig.~\ref{fig:004_peak}(b)], the image becomes strongly inhomogeneous. Regions of enhanced and reduced diffracted intensity indicate that different parts of the illuminated volume satisfy the Bragg condition at slightly different angular settings, as expected for remanent lattice rotations and stable extended defects that remain after the external stress is removed. The fractured sample [Fig.~\ref{fig:004_peak}(c)] shows still stronger contrast variations, including high-intensity bands and broad inhomogeneous regions, consistent with larger lattice rotations and a higher density of stress-induced extended defects generated during mechanical failure. A qualitative difference is also evident in the spatial character of the contrast. In the moderately compressed sample, the diffracting regions form a more distributed and only weakly directional pattern, whereas the fractured sample exhibits more pronounced band-like contrast with a preferred orientation. This difference is consistent with a redistribution and partial relaxation of the remanent deformation during unloading and fracture, although the peak images alone do not allow the underlying deformation mechanism to be identified. Because the color scale is adjusted independently for each panel, the important information in Fig.~\ref{fig:004_peak} is the spatial inhomogeneity within each image, not the absolute intensity difference between different samples.

The widths of the angular scans support the presence of increased lattice misorientation after compression. Representative widths obtained from the integrated diffracted intensity are $\Delta\mu=0.014^\circ$ for the pristine sample and $\Delta\mu=0.032^\circ$ for the moderately compressed sample, whereas the corresponding $\chi$ widths are approximately $0.18^\circ$ for both samples. The rocking and rolling scans have substantially different instrumental angular resolutions in DFXM, with the rolling response being considerably broader because of the reciprocal-space acceptance of the imaging optics~\cite{Poulsen2017}. The more than twofold increase of $\Delta\mu$ therefore provides clear evidence for an increased mosaic spread after compression, whereas the absence of a detectable change in $\Delta\chi$ should not be interpreted as evidence for the absence of additional misorientation in the orthogonal direction.

The susceptibility data in Fig.~\ref{fig:compression}(c) show that these changes in DFXM contrast are not accompanied by a measurable shift or broadening of the Verwey transition. Thus, the defects produced by uniaxial compression are clearly visible structurally, but their influence on the macroscopic transition temperature is weak compared with the effect of point-like disorder. This comparison is important because the DFXM images demonstrate that the absence of a change in $T_{\rm V}$ is not caused by an absence of deformation; rather, it shows that the particular defects produced here by uniaxial compression do not perturb the transition strongly enough to be detected in bulk ac susceptibility.

\subsection{Lattice misorientation and strain mapping}

Figure~\ref{fig:XRDMF} shows spatially resolved DFXM maps derived from the $(400)$ reflection for the same three samples. All maps were measured after unloading, i.e., without external stress applied during DFXM imaging. Therefore, the strain maps show residual axial strain fields in the unloaded samples, not the elastic strain present under applied compression. These maps provide the quantitative information that is only qualitatively visible in the peak images of Fig.~\ref{fig:004_peak}. The top row shows maps of the center of mass of the local $\mu$ rocking curves, which reveal spatial variations in the local lattice orientation. The bottom row shows strain maps obtained from shifts of the Bragg angle $2\theta$, which reflect local variations in lattice spacing along the scattering vector. The orientation and strain maps are complementary but do not exhibit a simple one-to-one spatial correspondence. Both show substantially stronger spatial variations after compression, indicating that the remanent deformation involves coupled lattice rotations and local strain fields, although the two quantities probe different components of the deformation. The strain values should therefore be understood as local axial strain components associated with the measured reflection, not as the full strain tensor.

In the pristine sample [Figs.~\ref{fig:XRDMF}(a) and \ref{fig:XRDMF}(d)], the mosaicity map is nearly uniform, and the axial strain map shows a comparatively homogeneous distribution of small positive and negative strain components within the illuminated area. This confirms the high crystalline quality of the starting material. In the sample compressed along $[011]$ to $200~\mathrm{MPa}$ [Figs.~\ref{fig:XRDMF}(b) and \ref{fig:XRDMF}(e)], diagonal line-like and band-like features appear in the local-orientation map, with local misorientations reaching approximately $\pm0.025^\circ$. At this stage of the analysis we identify these features by their trace direction, which is close to a $\langle 110\rangle$-type direction in the observed projection; their crystallographic origin is discussed below. The corresponding axial strain map displays a more pronounced heterogeneous pattern of local tensile and compressive regions. The stronger red and blue contrast indicates an increased amplitude of local strain variations, but not a clear uniform shift of the strain distribution over the illuminated area relative to the pristine sample.

In the fractured sample [Figs.~\ref{fig:XRDMF}(c) and \ref{fig:XRDMF}(f)], the local-orientation map shows larger-scale lattice rotations and fragmentation. This indicates that part of the crystal has rotated away from the orientation selected in the pristine sample, producing a broader distribution of local lattice orientations. The fractured sample does not show a simple monotonic increase of axial strain relative to the moderately compressed sample, because part of the accumulated elastic strain can be released during cracking. Instead, the fractured sample displays a redistributed residual strain pattern, including stripe-like contrast approximately along $[001]$, which may indicate a contribution from an additional slip-related deformation mode or redistribution of strain after crack formation. The coexistence of strong mosaicity contrast with a redistributed strain pattern is consistent with plastic deformation followed by partial elastic relaxation during fracture.

\subsection{Fracture of the sample compressed along $[001]$}

\begin{figure}[t]
\centering
\includegraphics[width=0.48\textwidth]{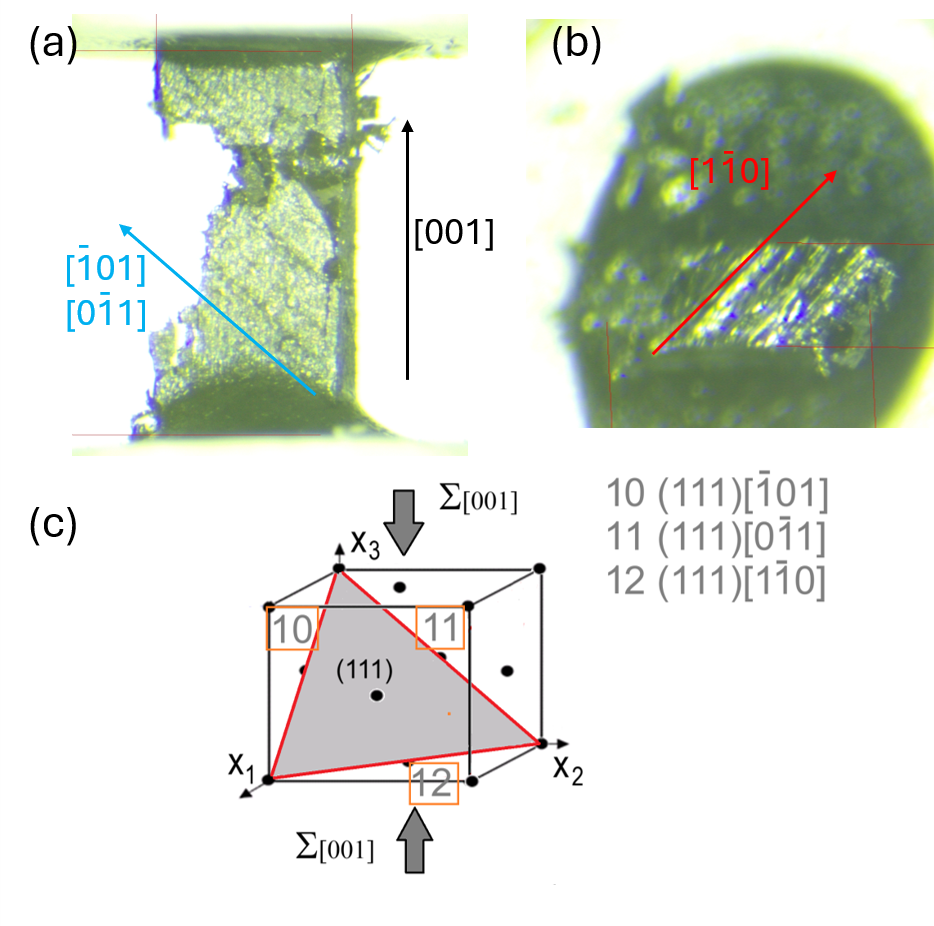}
\caption{Fracture morphology and slip traces of a magnetite single crystal compressed along the cubic $[001]$ direction.
(a) Side view of the fractured sample. The black arrow indicates the $[001]$ compression direction, and the blue arrow marks an observed oblique slip trace compatible with the projected $[\bar{1}01]$ or $[0\bar{1}1]$ directions.
(b) Top view of the same sample, showing an observed slip trace approximately along the $[1\bar{1}0]$ direction, marked by the red arrow.
(c) Schematic illustration of selected $(111)\langle 110\rangle$ slip systems, correlated with the corresponding slip traces observed on the specimen surface under uniaxial stress $\boldsymbol{\Sigma}_{[001]}$. The gray arrows indicate compression along $[001]$. The labels 10, 11, and 12 denote the slip systems $(111)[\bar{1}01]$, $(111)[0\bar{1}1]$, and $(111)[1\bar{1}0]$, respectively, used in the resolved-shear-stress analysis and listed in Table~\ref{tab:rss}. For ideal $\boldsymbol{\Sigma}_{[001]}$ loading, systems 10 and 11 have nonzero RSS, whereas system 12 has zero RSS and is included as a geometrical reference for the observed top-view trace.}
\label{fig:fracture}
\end{figure}

Optical microscopy images of the sample compressed along $[001]$ are shown in Fig.~\ref{fig:fracture}. The fracture pattern is not random. In the side view [Fig.~\ref{fig:fracture}(a)], the damage forms an oblique slip trace inclined relative to the compression axis and compatible with the projected $[\bar{1}01]$ or $[0\bar{1}1]$ directions. In the top view [Fig.~\ref{fig:fracture}(b)], the observed slip trace is aligned approximately along $[1\bar{1}0]$. This directional character shows that fracture is controlled by the crystallography of the spinel lattice rather than by arbitrary surface damage. The schematic in Fig.~\ref{fig:fracture}(c) indicates the $(111)$ plane and the $\langle 110\rangle$ directions relevant for discussing the selected slip systems. The labels 10, 11, and 12 refer to the corresponding slip systems listed in Table~\ref{tab:rss}. The relation between these fracture traces and the line-like defects visualized by DFXM is discussed below.

\subsection{Slip systems and resolved shear-stress calculations}
\label{sec:slip}

The fracture morphology shown in Fig.~\ref{fig:fracture} and the line-like contrast observed in the DFXM maps suggest that plastic deformation is crystallographically controlled. To compare the two loading geometries relevant to the experiment, we calculated the resolved shear stress (RSS) for the 12 standard fcc $\{111\}\langle 110\rangle$ slip systems associated with the oxygen sublattice of the spinel structure~\cite{Till2020}. Since the critical resolved shear stress was not determined for the present samples, the calculation is used only to compare the relative number of slip systems with nonzero RSS, not to predict an absolute yield condition.

For a slip system $s$, we denote the resolved shear stress by $\tau_s$. It is given by the projection of the stress tensor onto the slip direction and the slip-plane normal,
\begin{equation}
\tau_s =
m_i^s \Sigma^{\rm crs}_{ij} n_j^s .
\end{equation}
Here $\Sigma^{\rm crs}_{ij}$ is the Cauchy stress tensor expressed in the cubic crystallographic basis, $\mathbf{n}^s$ is the unit vector normal to the slip plane $(hkl)$, and $\mathbf{m}^s$ is the unit vector parallel to the slip direction $[uvw]$. The superscript ``crs'' denotes crystal coordinates and distinguishes this tensor from the applied uniaxial stress tensors $\boldsymbol{\Sigma}_{[001]}$ and $\boldsymbol{\Sigma}_{[0\bar{1}1]}$ used below. The indices $i,j=1,2,3$ denote Cartesian components along the cubic crystallographic axes used in the calculation, and repeated indices are summed over $i$ and $j$ according to the Einstein convention.

At the onset of slip, the absolute value of the resolved shear stress reaches the critical resolved shear stress, $\tau_{\rm cr}$,
\begin{equation}
|\tau_s| = \tau_{\rm cr}.
\end{equation}

\begin{figure*}[th]
\centering
\includegraphics[width=0.6\textwidth]{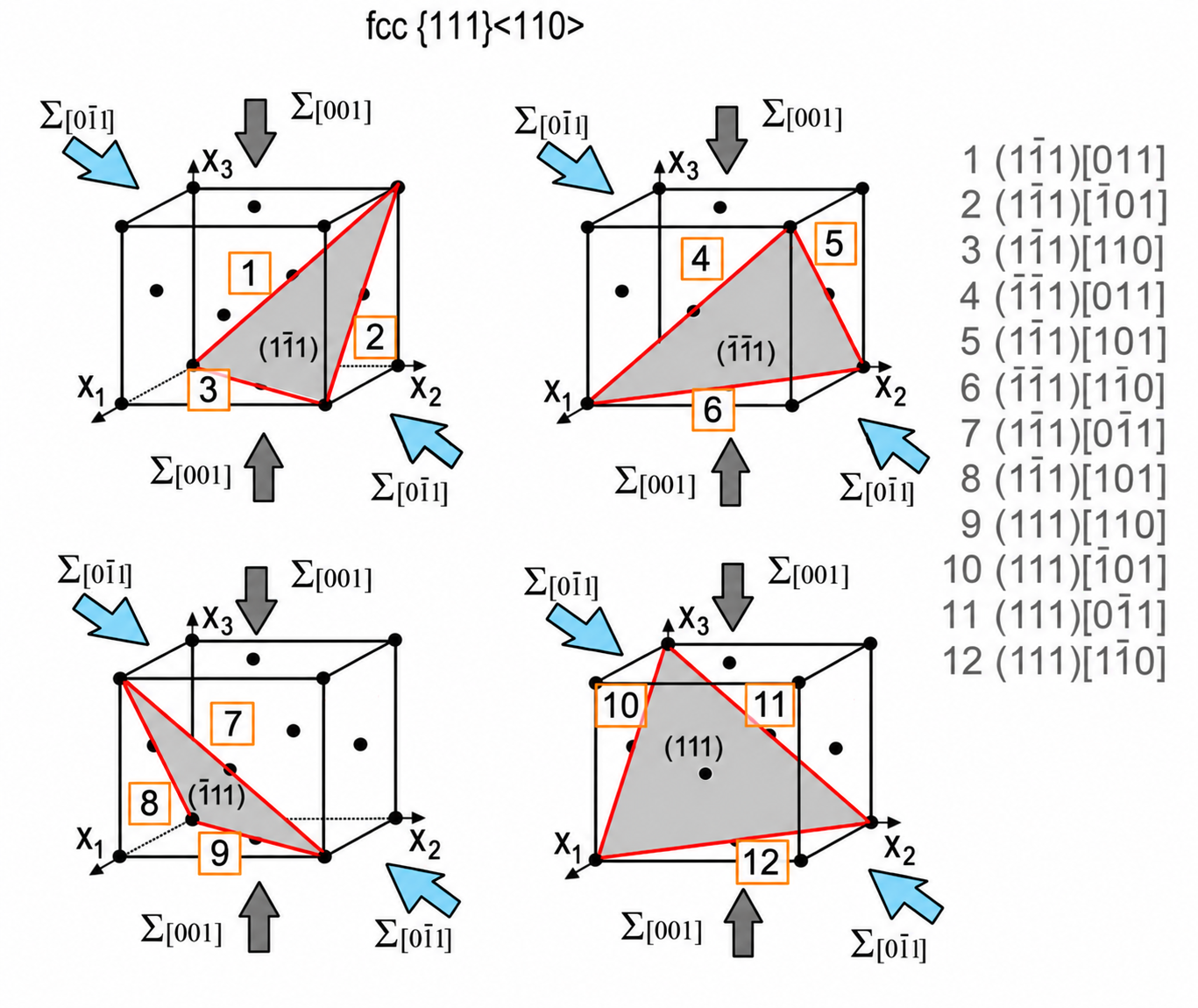}
\caption{Schematic numbering of the 12 fcc $\{111\}\langle 110\rangle$ slip systems used in the resolved-shear-stress calculation. The four panels show the four symmetry-equivalent $\{111\}$ slip planes, and the numbered labels denote the corresponding $\langle 110\rangle$ slip directions. Gray arrows indicate the uniaxial stress $\boldsymbol{\Sigma}_{[001]}$ applied along the $[001]$ direction. Blue arrows indicate the uniaxial stress $\boldsymbol{\Sigma}_{[0\bar{1}1]}$ applied along the $[0\bar{1}1]$ direction. The same numbering is used in Table~\ref{tab:rss}.}
\label{fig:slip_systems}
\end{figure*}

The calculation was performed in the cubic crystallographic reference frame, which is the notation used throughout this manuscript. Loading along $[001]$ represents the sample compressed along the $[001]$ direction. In the RSS calculation and in Fig.~\ref{fig:slip_systems}, the second loading geometry is written as $[0\bar{1}1]$, because this is the direction shown by the blue arrows in the schematic. This direction is symmetry-equivalent to the $[011]$ direction used for the DFXM samples. All symmetry-equivalent slip systems are considered. The stress tensors below correspond to a uniaxial compressive stress of $100~\mathrm{MPa}$ and are expressed in the cubic crystal frame.

\begin{table*}
\caption{Resolved shear stresses for the 12 slip systems of the $\{111\}\langle 110\rangle$ family under uniaxial stress $\boldsymbol{\Sigma}_{[001]}$ and $\boldsymbol{\Sigma}_{[0\bar{1}1]}$. Values are absolute RSS values in MPa for a uniaxial compressive stress of $100~\mathrm{MPa}$.}
\label{tab:rss}
\begin{ruledtabular}
\begin{tabular}{c|cccccccccccc}
Slip system no. & 1 & 2 & 3 & 4 & 5 & 6 & 7 & 8 & 9 & 10 & 11 & 12 \\
\hline
RSS, $\boldsymbol{\Sigma}_{[001]}$ &
40.8 & 40.8 & 0 & 40.8 & 40.8 & 0 & 40.8 & 40.8 & 0 & 40.8 & 40.8 & 0 \\
RSS, $\boldsymbol{\Sigma}_{[0\bar{1}1]}$ &
0 & 40.8 & 40.8 & 0 & 40.8 & 40.8 & 0 & 0 & 0 & 0 & 0 & 0
\end{tabular}
\end{ruledtabular}
\end{table*}

For uniaxial stress $\boldsymbol{\Sigma}_{[001]}$ applied along $[001]$, the stress tensor expressed in the cubic crystal frame is
\begin{equation}
\Sigma^{\rm crs}_{ij}([001]) =
\begin{pmatrix}
0 & 0 & 0 \\
0 & 0 & 0 \\
0 & 0 & -100
\end{pmatrix}
~\mathrm{MPa},
\end{equation}
while for uniaxial stress $\boldsymbol{\Sigma}_{[0\bar{1}1]}$ applied along $[0\bar{1}1]$, the stress tensor in the same frame is
\begin{equation}
\Sigma^{\rm crs}_{ij}([0\bar{1}1]) =
\begin{pmatrix}
0 & 0 & 0 \\
0 & -50 & 50 \\
0 & 50 & -50
\end{pmatrix}
~\mathrm{MPa}.
\end{equation}
The values in Table~\ref{tab:rss} are normalized to $100~\mathrm{MPa}$ and scale linearly with the applied stress. For a loading level of approximately $200~\mathrm{MPa}$, the absolute RSS values would therefore be twice those listed in the table. Since the analysis is used only to compare the two loading geometries, the relevant result is the number of systems with nonzero RSS, not the absolute RSS value.

Figure~\ref{fig:slip_systems} defines the numbering of the 12 $\{111\}\langle 110\rangle$ slip systems used in the calculation. Table~\ref{tab:rss} summarizes the corresponding absolute RSS values. The labels 10, 11, and 12 shown in Fig.~\ref{fig:fracture}(c) refer to the same slip-system numbering.

For uniaxial stress applied along $[001]$, which represents compression along the cubic $[001]$ direction, eight of the twelve $\{111\}\langle 110\rangle$ slip systems have nonzero RSS values. The exceptions are systems 3, 6, 9, and 12, corresponding to $(1\bar{1}1)[110]$, $(\bar{1}\bar{1}1)[1\bar{1}0]$, $(\bar{1}11)[110]$, and $(111)[1\bar{1}0]$, respectively. In all four cases, the slip direction has no $[001]$ component, and therefore the resolved shear stress vanishes for ideal $[001]$ uniaxial loading. In contrast, for stress applied along $[0\bar{1}1]$, only four slip systems have nonzero RSS values. In the ideal $[001]$ loading geometry considered here, systems 10 and 11 have nonzero RSS, whereas system 12 has zero RSS.

This difference in slip-system multiplicity provides a useful geometrical comparison of the two loading directions. Because the critical resolved shear stress and the fracture criterion were not determined for the present samples, the RSS calculation does not by itself predict which loading geometry should fracture more readily. Nevertheless, the larger number of $\{111\}\langle 110\rangle$ systems with nonzero RSS under $[001]$ loading is qualitatively consistent with the greater tendency to fracture observed in this geometry, whereas the symmetry-equivalent $[0\bar{1}1]$ geometry gives nonzero RSS for fewer systems.

The calculation therefore supports the interpretation of the DFXM and optical-microscopy results. The line-like defects observed in the compressed samples are consistent with slip-related deformation along $\{111\}$ planes and $\langle 110\rangle$-type directions, as shown by the projections in Fig.~\ref{fig:fracture}. In particular, the inclination of the damage traces in the optical images and the line-like contrast in the DFXM maps are consistent with the expected traces of slip-related deformation under uniaxial compression.

\section{Discussion}
\label{sec:discussion}

The DFXM measurements on the two samples compressed along $[011]$ demonstrate that uniaxial compression leaves a clear remanent structural signature in magnetite. Compared with the pristine crystal, the compressed samples exhibit line-like and band-like contrast in the $\mu$ rocking-curve maps and heterogeneous residual strain in the $2\theta$ maps. These features are absent or weak in the reference crystal and become pronounced after compression and fracture, showing that the applied stress produces persistent extended defects. The comparison is particularly direct because the samples were cut from the same parent crystal and were measured after unloading, so the observed differences reflect the remanent defect state rather than the instantaneous elastic response to an applied force.

The dominant line-like features run approximately along $\langle 110\rangle$-type directions in the observed projection. A similar conclusion follows from the optical image of the sample compressed along $[001]$, where the main fracture trace is approximately along $[1\bar{1}0]$. Such traces can in principle correspond to projections of more than one crystallographic plane family onto the observed surface, so the plane assignment cannot be made from the optical image alone. However, slip in spinel-structured magnetite is commonly discussed in terms of $\langle 110\rangle$-type directions and low-index planes of the oxygen framework~\cite{Till2020}. Room-temperature deformation studies of magnetite have also identified $\{100\}$ and $\{111\}$ as possible slip planes~\cite{Charpentier1968}. In the present geometry, the DFXM line-like defects and the fracture morphology are therefore most naturally interpreted as slip-related features associated mainly with $\{111\}$ planes. This assignment is also consistent with the slip-system analysis in Sec.~\ref{sec:slip}. The three systems drawn in Fig.~\ref{fig:fracture}(c) correspond to $(111)[\bar{1}01]$, $(111)[0\bar{1}1]$, and $(111)[1\bar{1}0]$, labeled 10, 11, and 12, respectively, in Table~\ref{tab:rss}. For ideal $\boldsymbol{\Sigma}_{[001]}$ loading, the $(111)[\bar{1}01]$ and $(111)[0\bar{1}1]$ systems have nonzero RSS, whereas $(111)[1\bar{1}0]$ has zero RSS and is therefore included as a geometrical reference for the observed top-view trace. The additional stripe-like contrast visible in the strain map of the fractured sample [Fig.~\ref{fig:XRDMF}(f)] may indicate a contribution from another slip system, possibly involving $\{100\}$ planes, or a redistribution of strain after cracking.

The anisotropic fracture behavior also supports this interpretation. The sample compressed along $[001]$ fractured readily, whereas samples loaded along $[011]$ could sustain larger deformation before failure. As shown by the resolved shear-stress analysis in Sec.~\ref{sec:slip}, loading along $[001]$ gives nonzero RSS for eight of the twelve $\{111\}\langle 110\rangle$ slip systems. By contrast, the symmetry-equivalent $[0\bar{1}1]$ loading geometry used in the RSS calculation gives nonzero RSS for only four systems. This difference in slip-system multiplicity is consistent with easier fracture for compression along $[001]$, whereas the experimental $[011]$ compression geometry, represented in the RSS calculation by the symmetry-equivalent $[0\bar{1}1]$ direction, more readily produces remanent line-like defects without immediate catastrophic failure.

The central result is that these remanent structural defects do not measurably affect the Verwey transition. The susceptibility curves in Fig.~\ref{fig:compression}(c) show that $T_{\rm V}$ and the transition sharpness remain essentially unchanged even after fracture. This is in strong contrast to point defects introduced by electron irradiation~\cite{Walz1990,Prozorov2023}, and to nonstoichiometry or doping, all of which suppress $T_{\rm V}$. The comparison suggests that the microscopic character of disorder is decisive: point-like defects couple efficiently to the electronic ordering involved in the Verwey transition, whereas extended slip-related defects and remanent strain fields produced by uniaxial compression do not. This does not imply that local strain near a dislocation is irrelevant at the microscopic scale. Rather, the volume-averaged susceptibility shows that the density and spatial distribution of defects generated in the present experiments are insufficient to shift or broaden the macroscopic transition within our experimental resolution.

This conclusion also provides a useful perspective on pressure effects. Hydrostatic pressure lowers $T_{\rm V}$ while it is applied, whereas pressure cycling into the GPa range can leave a persistent increase in $T_{\rm V}$ after decompression, with the magnitude depending on peak pressure and sample composition~\cite{Caporzen2010,Bialo2019}. Our results do not contradict these pressure-cycling studies, because the stress geometry, pressure range, and type of remanent disorder generated by uniaxial plastic deformation need not be the same as those produced by high-pressure cycling. The present DFXM results show that a substantial remanent strain field is not sufficient by itself to shift $T_{\rm V}$ if it is mainly associated with extended line-like defects. Therefore, the pressure memory of $T_{\rm V}$ likely involves a different microscopic defect population, a different structural rearrangement, or defect configurations that couple more directly to the charge-ordering pattern than the slip-related defects generated here.

Overall, the experiments distinguish two kinds of lattice disorder in magnetite. Point-like disorder, produced by irradiation, nonstoichiometry, or chemical substitution, strongly perturbs the Verwey transition. Extended disorder, produced here by plastic deformation and preserved as slip-related line defects and residual strain fields, produces strong structural contrast but leaves the transition unchanged within experimental resolution. This distinction is essential for interpreting strain and defect effects in magnetite. It also emphasizes that a broadened or shifted Verwey anomaly should not automatically be attributed to generic lattice damage or residual strain. Instead, the relevant question is whether the defect structure changes the local electronic balance and the charge-ordering network. This point may also apply to other correlated oxides in which electronic order coexists with ferroelastic or magnetostructural degrees of freedom.

\section{Summary and conclusions}

We used DFXM and ac magnetic susceptibility to study the effect of uniaxial compression on the lattice microstructure and the Verwey transition in stoichiometric single-crystal magnetite. Compression to $200~\mathrm{MPa}$ along $[011]$ produced stable line-like defects and remanent strain fields that persisted after unloading. Stronger compression along $[011]$ caused fracture and a denser defect network, whereas compression along $[001]$ led to catastrophic fracture. The observed line-like features and fracture traces are consistent with slip-related defects, mainly associated with $\{111\}\langle 110\rangle$ slip systems, as indicated by the resolved shear-stress analysis.

Despite these pronounced microstructural changes, no measurable shift or broadening of $T_{\rm V}$ was observed. This result contrasts with the known suppression of $T_{\rm V}$ by point defects, nonstoichiometry, and doping. We therefore conclude that extended slip-related defects and remanent strain fields produced by uniaxial compression do not couple strongly to the electronic ordering responsible for the Verwey transition, whereas point-like disorder perturbs it much more effectively.

More generally, our results and literature data show that not all lattice defects are equivalent in correlated oxides. In magnetite, plastic deformation can produce substantial remanent strain and extended line-like defects without changing the Verwey transition, providing a clear separation between structural damage and the electronic instability that drives the transition.

\begin{acknowledgments}

This work was cofinanced by the National Science Centre (NCN), Poland, under OPUS grant no. 2021/41/B/ST3/03454. A.B. was supported by the NCN under grant no. 2023/49/B/ST11/00774. The research was partly supported by the ``Excellence Initiative -- Research University'' program for the AGH University of Krakow, Project No. 6387. N.B. was supported by the Croatian Science Foundation under Project No. IP-2022-10-3382 and by the CeNIKS project, cofinanced by the Croatian Government and the European Union through the European Regional Development Fund, Competitiveness and Cohesion Operational Program, Grant No. KK.01.1.1.02.0013. The work at TU Wien was supported by the Austrian Science Fund (FWF) [10.55776/F86; 10.55776/P35945]. The DFXM experiments were performed at the European Synchrotron Radiation Facility (ESRF), Grenoble, France, under proposals HC-4565 and HC-5587. The access to ESRF was financed by the Polish Ministry of Science and Higher Education, decision no. 2021/WK/11.

\end{acknowledgments}

\newpage

\bibliography{MagnetitBib}

\end{document}